\documentclass[sigconf, anonymous=False]{acmart}
\renewcommand\footnotetextcopyrightpermission[1]{} 
\usepackage{tabularx, booktabs}
\usepackage{graphicx, multicol, footmisc}
\usepackage{multirow}
\usepackage{bm}
\usepackage{balance}
\usepackage{enumitem}
\usepackage{amsmath}
\usepackage{diagbox}
\usepackage[caption=false]{subfig}
\usepackage{siunitx}
\usepackage{adjustbox}
\usepackage{array}

\newcolumntype{x}[1]{>{\centering\arraybackslash\hspace{0pt}}p{#1}}

\newcolumntype{C}[1]{>{\centering\arraybackslash}p{#1}}
\newcolumntype{R}[1]{>{\raggedleft\arraybackslash}p{#1}}
\newcolumntype{L}[1]{>{\raggedright\arraybackslash}p{#1}}

\newcommand\mc[1]{\multicolumn{1}{c}{#1}}
\newcommand\mcl[1]{\multicolumn{1}{c|}{#1}}

\newcommand{\boldentry}[2]{%
  \multicolumn{1}{S[table-format=#1,
                    mode=text,
                    text-rm=\fontseries{b}\selectfont
                   ]}{#2}}

\newcommand{\boldentryline}[2]{%
  \multicolumn{1}{S[table-format=#1,
                    mode=text,
                    text-rm=\fontseries{b}\selectfont
                   ]|}{#2}}

\newcommand\blfootnote[1]{%
\begingroup 
\renewcommand\thefootnote{}\footnote{#1}%
\addtocounter{footnote}{-1}%
\endgroup 
}

\begin{document}
\fancyhead{}

\title{Interpreting Dense Retrieval as Mixture of Topics}

\author{Jingtao Zhan$^{1}$, Jiaxin Mao$^{2}$, Yiqun Liu$^{1\star}$, Jiafeng Guo$^{3}$, Min Zhang$^{1}$, Shaoping Ma$^{1}$}
\affiliation{%
  \institution{${1}$ Department of Computer Science and Technology, Institute for Artificial Intelligence, \\
  Beijing National Research Center for Information Science and Technology, Tsinghua University, Beijing 100084, China
  }
  }
  
\affiliation{%
  \institution{${2}$ Beijing Key Laboratory of Big Data Management and Analysis Methods, Gaoling School of Artificial Intelligence, \\ Renmin University of China, Beijing 100872, China
}
  }

\affiliation{%
  \institution{${3}$ CAS Key Lab of Network Data Science and Technology, Institute of Computing Technology, \\
  		Chinese Academy of Sciences, Beijing, China
  }
  }

\email{jingtaozhan@gmail.com, maojiaxin@gmail.com, yiqunliu@tsinghua.edu.cn, guojiafeng@ict.ac.cn}

\renewcommand{\shortauthors}{Zhan, et al.}

\begin{abstract}
Dense Retrieval~(DR) reaches state-of-the-art results in first-stage retrieval, but little is known about the mechanisms that contribute to its success. Therefore, in this work, we conduct an interpretation study of recently proposed DR models. Specifically, we first discretize the embeddings output by the document and query encoders. Based on the discrete representations, we analyze the attribution of input tokens. Both qualitative and quantitative experiments are carried out on public test collections. Results suggest that DR models pay attention to different aspects of input and extract various high-level topic representations. Therefore, we can regard the representations learned by DR models as a mixture of high-level topics.   
\end{abstract}

\maketitle

\blfootnote{$^\star$Corresponding author}

\section{Introduction}

Dense Retrieval~(DR) achieves state-of-the-art first-stage retrieval performance through representing text with dense embeddings and utilizing (approximate) nearest neighbor search to retrieve candidates~\cite{karpukhin2020dense,zhan2021optimizing,zhan2021learningdr,hofstatter2021efficiently,gao2020complementing,luan2020sparsedense}. However, the mechanisms that contribute to its outstanding performance remain unclear.

We address this problem by conducting an interpretation study. Firstly, we propose RepMoT, which discretizes the output of query and document encoders. The discrete representations make interpretation study easier by modeling text relationships more explicitly than continuous embeddings. Secondly, we investigate DR models based on RepMoT with attribution techniques. We qualitatively analyze the word importance with respect to different discrete sub-vectors. Then we quantitatively measure the word importance by masking the input and observing the output change. 

We conduct experiments on two widely-adopted benchmark datasets. Experimental results show RepMoT achieves ranking performance on par with state-of-the-art DR models and thus justifies analyzing DR models based on RepMoT. The analysis results suggest DR models pay attention to different aspects of the input and capture mixture of high-level topics. Therefore, the representations encoded by DR models can be regarded as mixture of high-level topic representations.

\section{Related Works}


DR models embed queries and documents to low-dimension dense vectors and utilize their vector similarity to predict relevance~\cite{zhan2020repbert,gao2021unsupervised,zhan2021optimizing,huang2020embedding,hofstatter2021efficiently}. Formally, let $q$ be the user query and $d$ be the document. We use $f$ to denote the DR model. Thus, $\bm{q} = f(q) \in \mathbb{R}^{D}$ and $ \bm{d} = f(d) \in \mathbb{R}^{D}$, where $D$ is the embedding dimension.

Previous DR models usually embed text to a continuous embedding space, such as ANCE~\cite{xiong2021approximate} and ADORE~\cite{zhan2021optimizing}. Recently, several studies jointly optimize DR models with Product Quantization~\cite{jegou2010product} and produce discrete document representations, such as JPQ~\cite{zhan2021jointly} and RepCONC~\cite{zhan2021learningdr}. Our proposed RepMoT further discretizes query representations based on RepCONC~\cite{zhan2021learningdr}.

Here is a brief note about the term ``Topic''. ``Topic'' in the context of topic models usually refers to clusters of similar words and is mined based on word statistics~\cite{blei2012probabilistic}. In this paper, the term ``Topic'' is used in its idiomatic sense, and we investigate which words DR models pay attention to. Future works may further explore the relationships between RepMoT and topic models. 



\section{Methodology}

\subsection{Discrete Representations}

It is difficult to directly interpret the DR models in the continuous embedding space because we need to manually set a threshold for the \emph{soft} matching score to determine whether two pieces of texts are semantically similar or not. Inspired by Product Quantization~\cite{jegou2010product}, we propose to discretize the query and document representations. The discrete representations explicitly characterize the relationship between different texts. The queries or documents assigned to the same discrete sub-vectors are similar and those to different ones are dissimilar. 


\begin{table*}[t]
    \centering
    \caption{
    Retrieval results on TREC 2019 Deep Learning Track. M denotes the number of discrete sub-vectors per passage/document. 
    }
    \scalebox{0.95}{
    \begin{tabular}{@{}ll|S[table-column-width=10mm]S[table-column-width=7mm]S[table-column-width=7mm]|
    S[table-format=1.3,table-column-width=16mm]S[table-format=1.3,table-column-width=16mm]S[table-format=1.3,table-column-width=16mm]S[table-format=1.3,table-column-width=16mm]
    |
    S[table-format=1.3,table-column-width=16mm]S[table-format=1.3,table-column-width=16mm]@{}}
    \toprule
    \textbf{} & \multirow{2}{*}{\textbf{Model}} & \multicolumn{3}{c|}{\textbf{Discrete}} & \multicolumn{2}{c}{\textbf{MARCO Passage}} & \multicolumn{2}{c|}{\textbf{DL Passage}} & \multicolumn{2}{c}{\textbf{MARCO Doc}} \\
     &  & {Query} & {Doc} & {M} & {MRR@10} & {R@100} & {NDCG@10} & {R@100}  & {MRR@100} & {R@100} \\ \midrule
     & ANCE~\cite{xiong2021approximate} & & & {$\backslash$} & 0.330 & 0.852 & 0.648 & 0.437 & 0.377 & 0.894 \\ 
     & ADORE~\cite{zhan2021optimizing} & & & {$\backslash$} & 0.347 & 0.876 & 0.683 & 0.473 & 0.405 & 0.919 \\ \midrule
     & OPQ~\cite{jegou2010product} &  & {\checkmark} & {48} & 0.290 & 0.830 & 0.591 & 0.417 & 0.340 & 0.880 \\
     & JPQ~\cite{zhan2021jointly} &  & {\checkmark} & {48} & 0.332 & 0.863 & 0.644 & 0.447 & 0.384 & 0.905 \\ 
     & RepCONC~\cite{zhan2021learningdr} &  & {\checkmark} & {48} & 0.340 & 0.864 & 0.668 & 0.492 & 0.399 & 0.911 \\ \midrule
     & OPQ-SDC~\cite{jegou2010product} & {\checkmark} & {\checkmark} & {48} & 0.285 & 0.822 & 0.592 & 0.404 & 0.338 & 0.873 \\
     & RepMoT & {\checkmark} & {\checkmark} & {48} & 0.335 & 0.854 & 0.661 & 0.441 & 0.394 & 0.905 \\
    \bottomrule
    \end{tabular}
    }
    \label{tab:discrete_rank}
    \end{table*}

Now we describe how to discretize continuous representations.
1) We first split the output of DR models to $M$ sub-vectors of dimension $D/M$. We use $f_i({\rm text}) \in \mathbb{R}^{D/M}$ to denote the $i$-th sub-vector. 
2) Next, we define $M$ embedding pools, each of which involves $K$ embeddings of dimension $D/M$. We use $\bm{c}_{i,j} \in \mathbb{R}^\frac{D}{M} $ to denote the $j$-th embedding from the $i$-th pool. 
3) Finally, we replace each sub-vector $f_i({\rm text})$ with the nearest embedding in the corresponding pool $\{\bm{c}_{i,j}\}$. Let $\bm{\hat{d}}$ be the discrete document representation. It is the concatenation of $M$ sub-vectors $\bm{\hat{d}_i}$:
\begin{equation}
\left\{
  \begin{array}{clr}
      \bm{\hat{d}} &= \bm{\hat{d}}_1,\bm{\hat{d}}_2,...,\bm{\hat{d}}_M\\
      \bm{\hat{d}}_i &= \bm{c}_{i,\varphi_i(d)} \text{ and }
\varphi_i(d) = \arg \min_j || f_i(d) - \bm{c}_{i,j} ||
  \end{array}
\right.
\label{eq:repmot_discretize}
\end{equation}
where commas denote concatenation. Query representations are defined in the same way. Each embedding pool splits the corpus into $K$ groups, and the texts in the same group are considered similar. The discrete representation is still very expressive because the total number of combinations is $K^M$.

Inspired by RepCONC~\cite{zhan2021learningdr}, we end-to-end optimize the dual-encoders $f$ and embedding pools $\{\bm{c}_{i,j}\}$ with the weighted sum of a ranking-oriented loss and a clustering loss. The ranking-oriented loss replaces the continuous embeddings in the common ranking loss functions~\cite{qu2021rocketqa,xiong2021approximate,zhan2021optimizing} with the discrete embeddings: 
\begin{equation}
L_{r} = - \log \frac{\mathrm{e}^{\langle \bm{\hat{q}}, \bm{\hat{d}}^+ \rangle}}{\mathrm{e}^{\langle \bm{\hat{q}}, \bm{\hat{d}}^+ \rangle} + \sum_{d^-} \mathrm{e}^{\langle \bm{\hat{q}}, \bm{\hat{d}}^- \rangle}} \\
\end{equation}
where $\hat{q}$ and $\hat{d}$ are discrete query and document representations. 
The clustering loss is the MSE loss between the discrete and continuous embeddings:
\begin{equation}
L_{m} = \frac{1}{2} (\lVert f(d) - \bm{\hat{d}} \rVert ^2 + \lVert f(q) - \bm{\hat{q}} \rVert ^2)
\end{equation}
We utilize their weighed sum $L_{r} + \lambda L_{m}$ as the final loss function. 
We also adopt the uniform clustering constraint~\cite{zhan2021learningdr} when discretizing document embeddings during training, which, according to RepCONC~\cite{zhan2021learningdr}, helps maximize the distinguishability of discrete representations. We refer readers to RepCONC~\cite{zhan2021learningdr} for details. 

We name our method RepMoT because the following sections will show it captures \textbf{M}ixture \textbf{o}f high-level \textbf{T}opics.

\subsection{Attribution Analysis}

We aim to interpret DR on word level by quantifying the contributions of each word to each discrete sub-vector by a scalar score. For example, say the DR model takes the input of the sentence ``\textit{his food is tasty.}". ``\textit{food}" and ``\textit{tasty}" are likely to have high contribution scores because they are important for the semantic meanings. ``\textit{is}" carries little information and thus has low score. 

We adopt a popular attribution method, Integrated Gradients (IG)~\cite{sundararajan2017axiomatic}, to measure the word importance. It interpolates the input features from the baseline input to the actual input and computes the partial derivatives of the model output with respect to the input. 
Formally, let $x = (x_1,...,x_n) \in \mathbb{R}^{n\times dim}$ be the actual input word embeddings, and $x' = (x'_1,...,x'_n) \in \mathbb{R}^{n\times dim}$ be the baseline input. The baseline can be a sequence of ``\textlangle{}\textit{unk}\textrangle{}" tokens
\footnote{
According to Zhan et al.~\cite{zhan2020analysis}, removing stop words leads to unexpected behaviors of BERT. Thus, we keep all stop words for the baseline input.
}
. Let $F$ be a neural model. The attribution of $j$-th word equals to:
\begin{equation}
(x_j-x_j') \cdot \int_{\alpha=0}^{1} \frac{\partial F(x'+\alpha (x-x'))}{\partial x_j} d \alpha
\end{equation}

Recall that RepMoT encodes the input text to the concatenation of multiple discrete sub-vectors. To study whether different discrete sub-vectors pay attention to different words, we investigate the attribution scores with respect to each output sub-vector. We use the following formula to compute the $j$-th word's attribution for $i$-th output sub-vector. 
\begin{equation}
(x_j-x_j') \cdot \int_{\alpha=0}^{1} \frac{\partial -||\bm{\hat{d}_i} - f_i(x'+\alpha (x-x')||^2}{\partial x_j} d \alpha
\label{eq:our_ig}
\end{equation}
The above attribution scores evaluate which words contribute to the selection of the discrete sub-vectors $\bm{\hat{d}_i}$. 
We do not directly use the output of $f_i$ but instead utilize the negative MSE. It is because as shown in Eq.~(\ref{eq:repmot_discretize}), MSE is important for selecting discrete representations while the exact output of DR models is not. In other words, $f_i(x'+\alpha (x-x'))$ is equivalent to $f_i(x)$ as long as it is closer to $\bm{\hat{d}_i}$ than to other embeddings from the pool. To model this characteristic, we use the MSE to compute gradients in Eq.~(\ref{eq:our_ig}). 

We compare the attribution distribution of the same text for different $\bm{\hat{d}_i}: 1\leq i \leq M $, and also analyze the highly attributed words of different text for the same $\bm{\hat{d}_i}$. Note, this paper focuses on analyzing document representations and leaves investigating query representations to future research.

\section{Experiments}

\begin{figure*}
	\subfloat[Energy]{\label{}\includegraphics[width=0.22\linewidth]{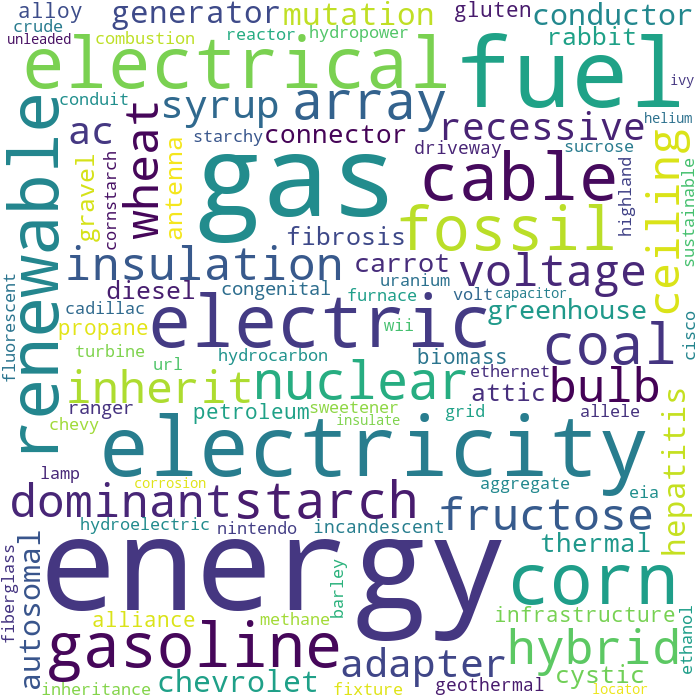}}  \hfill 
    \subfloat[Travel]{\label{}\includegraphics[width=0.22\linewidth]{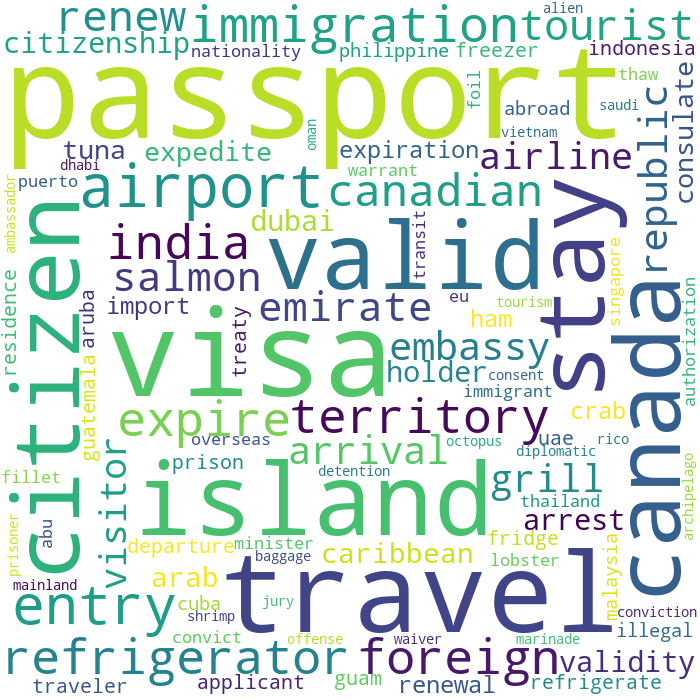}}  \hfill 
    \subfloat[Art]{\label{}\includegraphics[width=0.22\linewidth]{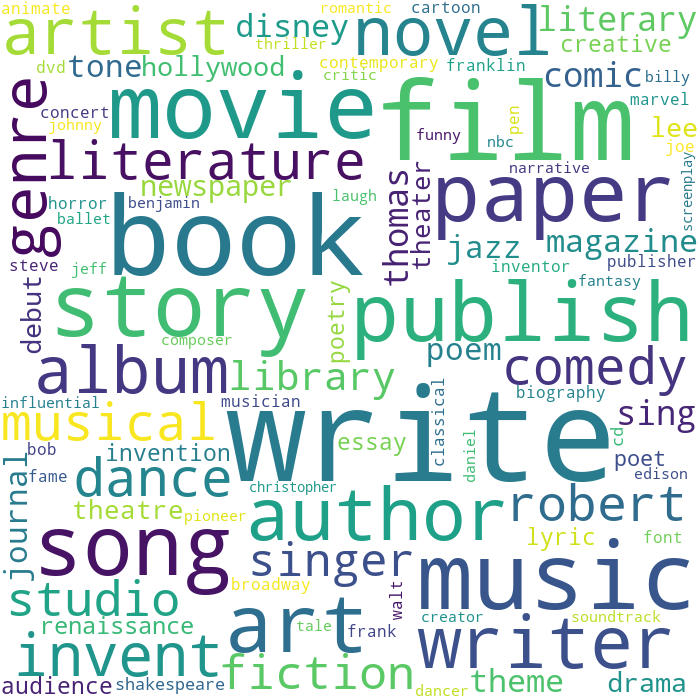}} \hfill 
    \subfloat[Geography]{\label{}\includegraphics[width=0.22\linewidth]{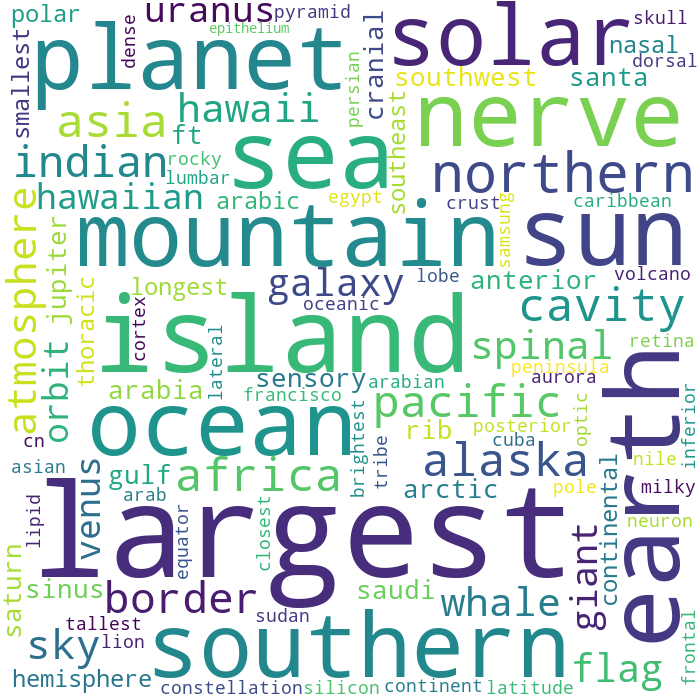}}
    \caption{Word cloud cases based on RepMoT on MS MARCO Passage Ranking. }
    \label{fig:wordcloud}
\end{figure*}

\begin{table*}[t]
    \centering
    \small
    \caption{The attribution results of one passage from MS MARCO Passage for different discrete embeddings. Topic-related words are manually selected from the highly attributed words of other passages assigned to the same discrete sub-vectors. }
    \scalebox{0.9}{
    \begin{tabular}{l | c} 
    \toprule
    Passage (highly attributed words are marked bold.)  & Topic-Related Words \\
    \midrule
    \begin{tabular}[l]{p{0.9\textwidth}} The \textbf{law} \textbf{defines} `full \textbf{time}" as 30 \textbf{hours} a \textbf{week} or more. ``We work so hard for so little pay," he said. ... This \textbf{month}, the administration delayed the employer insurance requirement until January \textbf{2015}.  \end{tabular} 
    &  \begin{tabular}[c]{@{}c@{}} day, week, \\ period, month \end{tabular}                  \\
    \midrule
    \begin{tabular}[l]{p{0.9\textwidth}} The \textbf{law} \textbf{defines} ``full time" as 30 hours a week or more. ``We work so hard for so little pay," he said. ... This month, the administration delayed the employer \textbf{insurance requirement} until January 2015.  \end{tabular} 
    &  \begin{tabular}[c]{@{}c@{}} government, health, \\ duty, insurance \end{tabular}                  \\
    \midrule
    \begin{tabular}[l]{p{0.9\textwidth}} The \textbf{law} \textbf{defines} ``full time" as 30 \textbf{hours} a week or more. ``We \textbf{work} so hard for so little pay," he said. ... This month, the administration delayed the employer insurance requirement until January 2015.  \end{tabular} 
    &  \begin{tabular}[c]{@{}c@{}} salary, wage, \\ work, pay \end{tabular}                  \\
    \bottomrule
    \end{tabular}
    }
    \label{tab:attr_case_study}
    \end{table*}

This section presents our experimental results. 
We conduct experiments on TREC 2019 Deep Learning Track~\cite{craswell2020overview,bajaj2016ms}. The passage ranking task has a corpus of $8.8$M passages, $0.5M$ training queries, $7k$ development queries~(henceforth, MARCO Passage), and $43$ test queries~(DL Passage). The document ranking task has a corpus of $3.2M$ documents, $0.4M$ training queries, $5k$ development queries~(MARCO Doc). 

The following sections first justify our analysis based on RepMoT, and then show DR models capture high-level topics. 

\subsection{Discrete Representations}

This section presents the ranking performance of RepMoT. 
We first introduce baselines and then discuss the experimental results. 

\subsubsection{Baselines} 
For continuous representations, we use ANCE~\cite{xiong2021approximate} and ADORE~\cite{zhan2021optimizing} as baselines. For half-discrete representations, we employ OPQ~\cite{ge2013optimized}, JPQ~\cite{zhan2021jointly}, and RepCONC~\cite{zhan2021learningdr}, all of which use discrete document representations and continuous query representations. For discrete representations, we adopt OPQ-SDC~\cite{ge2013optimized}, which unsupervisedly discretize representations with k-means. 

\subsubsection{Results}
Table~\ref{tab:discrete_rank} shows the experimental results. RepMoT achieves performance on par with the state-of-the-art DR models. It outperforms ANCE~\cite{xiong2021approximate} which employs continuous representations. It closes the gap with state-of-the-art ADORE~\cite{zhan2021optimizing} and RepCONC~\cite{zhan2021learningdr}, both of which employ (half-) continuous representations. Therefore, discrete representations learned by RepMoT are good approximations for DR models.

\subsection{Qualitative Analysis}

We first investigate the passages assigned to the same discrete sub-vector. Figure~\ref{fig:wordcloud} plots the word cloud examples for four sub-vectors. Clearly, the discrete sub-vector relates to a certain topic, e.g., energy, art, etc. Therefore, we can regard the discrete sub-vectors as topic representations. 

Next, we investigate which words RepMoT pays attention to when it encodes the text to multiple discrete sub-vectors. 
Table~\ref{tab:attr_case_study} visualizes the important words for three example discrete sub-vectors. The topic-related words in Table~\ref{tab:attr_case_study} are highly-attributed words of other passages assigned to the same sub-vectors.  
Results suggest that RepMoT pays attention to different aspects of the input to extract different high-level topic representations. Therefore, we can regard the learned representations as mixture of topics. 

\subsection{Quantitative Analysis}

We conduct a quantitative experiment to validate that different sub-vectors pay attention to different parts of the input. 
We mask certain words according to different assumptions and measure the following MSE: $||\bm{\hat{d}_i} - f_i(\rm{mask}(d))||$. We report the average MSE for different $i$. 
Smaller MSE indicates the unmasked words are important for $\bm{\hat{d}_i}$. 

According to our finding that different sub-vectors pay attention to different parts of the input, we design a masking method called MoT. For different $\bm{\hat{d}_i}$, MoT computes the attribution scores with Eq.~(\ref{eq:our_ig}) and thus can identify the important words specifically for a certain $\bm{\hat{d}_i}$. 
It keeps the top attributed words and masks others.
In the following, we first describe the baselines and then present the experimental results. 

\subsubsection{Baselines} 
All baselines assume the DR models pay attention to the same part of input for different sub-vectors.
Position-based methods include ``Tail", ``Head", and ``Rand" which only keep the tail, head, and random words while others are masked, respectively. Frequency-based methods involve ``IDF", ``TF", and "TF-IDF". Attribution-based methods involve ``GlobalT" and ``RandT". ``GlobalT" computes the global attribution scores with respect to $\bm{\hat{d}}$ instead of $\bm{\hat{d}_i}$. ``RandT" computes the attribution scores for another random discrete sub-vector $\bm{\hat{d}_j}$. 

\begin{table*}[t]
    \centering
    \small
    \caption{
    MSE error when only $5\%$ words are retained. ``12", ``24", and ``48" denote the number of sub-vectors. * denotes MoT performs significantly better than all baselines at $p<0.01$ level using two-tailed pairwise t-test. 
    Best method is marked bold. 
    }
    \scalebox{0.95}{
    \begin{tabular}{ll|
        S[table-format=1.2, table-column-width=12mm]
        S[table-format=1.2, table-column-width=12mm]
        S[table-format=1.2, table-column-width=12mm]
        |
        S[table-format=1.2, table-column-width=12mm]
        S[table-format=1.2, table-column-width=12mm]
        S[table-format=1.2, table-column-width=12mm]
        }
    \toprule
    \multicolumn{2}{l|}{\multirow{2}{*}{\diagbox[]{\textbf{Method}}{\textbf{Setup}}}} & \multicolumn{3}{c|}{\textbf{MARCO Passage}} & \multicolumn{3}{c}{\textbf{MARCO Document}}  \\ 
    & & \mc{\textbf{12}} & \mc{\textbf{24}} & \mcl{\textbf{48}} & \mc{\textbf{12}} & \mc{\textbf{24}} & \mc{\textbf{48}}  \\ \midrule
    \multicolumn{2}{l}{\textbf{Position-based}} \\
    & Tail 			& 4.20 & 2.66 & 1.82 & 3.06 & 2.24 & 1.14 \\ 
    & Rand 			& 4.18 & 2.64 & 1.80 & 3.22 & 2.27 & 1.16 \\ 
    & Head 			& 3.52 & 2.31 & 1.60 & 2.09 & 1.48 & 0.82 \\ 
    \multicolumn{2}{l}{\textbf{Frequency-based}} \\
    & IDF 			& 4.17 & 2.65 & 1.82 & 3.40 & 2.40 & 1.23 \\ 
    & TF 			& 3.63 & 2.32 & 1.58 & 2.01 & 1.49 & 0.80 \\ 
    & TF-IDF 		& 3.46 & 2.24 & 1.54 & 1.96 & 1.45 & 0.78 \\ 
    \multicolumn{2}{l}{\textbf{Attribution-based}} \\
    & RandT 		& 3.36 & 2.25 & 1.59 & 1.90 & 1.45 & 0.81 \\ 
    & GlobalT 		& 3.08 & 2.07 & 1.45 & 1.50 & 1.15 & 0.64 \\ 
    & MoT 			& \boldentry{1.2}{2.73*} & \boldentry{1.2}{1.87*} & \boldentryline{1.2}{1.26*} & \boldentry{1.2}{1.39*} & \boldentry{1.2}{1.05*} & \boldentry{1.2}{0.54*}\\ 
    \bottomrule
    \end{tabular}
    }
    \label{tab:mask_res}
    \end{table*}

\subsubsection{Results} Table~\ref{tab:mask_res} presents the experimental results.
It clearly shows that MoT performs significantly better than all masking baselines, demonstrating the word importance is different for different discrete sub-vectors. Therefore, we validate the finding in our qualitative analysis that DR models pay attention to different aspects of input to extract different high-level representations. 

\section{Conclusion and Future Work}

Recently, DR models yield strong first-stage retrieval performance, but the internal mechanisms leading to its success remain unknown. Therefore, we propose RepMoT to discretize embeddings and then analyze DR models based on RepMoT. Results suggest DR models pay attention to different aspects of the input and capture different high-level topics. Therefore, we regard the representations encoded by DR models as mixture of topics. 

We highlight several promising directions for building explainable DR. One is based on RepMoT, which can be regarded as re-writing the text with ``discrete vector terms''. Researchers may manually design the connections between those ``discrete vector terms'' and ``natural language terms'' during training. The other is teaching DR models how to capture topics by designing attention patterns with some heuristics, e.g., topic models. We leave these to future research.

%

\bibliographystyle{ACM-Reference-Format}
\balance
\bibliography{references}

\end{document}